\documentstyle[epsfig,indentfirst,bm,fleqn,12pt]{article}
\begin{document}
\begin{center}
{\large \bf  Further study on $\mathbf{5\textit{q}}$ configuration
states in the chiral SU(3) quark model \footnote[2]{Project
supported by the National Natural Science Foundation of China (No.
10475087).}}

\vspace{0.3cm}

{D. Zhang$^{a\footnotemark[3],b}$, F.
Huang$^{c,a\footnotemark[3],b}$, Z.Y. Zhang$^a$, Y.W. Yu$^a$}

\vspace{0.2cm}

{\small a) \it Institute of High Energy Physics, P.O. Box 918-4,
Beijing 100049, PR China}

{\small b) \it Graduate School of the Chinese Academy of Sciences,
Beijing 100049, PR China}

{\small c) \it CCAST (World Laboratory), P.O. Box 8730, Beijing
100080, PR China}
\end{center}

\footnotetext[3]{Mailing address.}

\vspace{0.3cm}

\begin{abstract}
The structure of the $5q$ configuration states with strangeness
${\cal{S}}=+1$ is further studied in the chiral SU(3) quark model
based on our previous work. We calculate the energies of fifteen
low configurations of the $5q$ system, four lowest configurations
of $J^{\pi}=\frac{1}{2}^-$ with $4q$ partition
$[4]_{orb}(0s^4)[31]^{\sigma f}$, four of $J^{\pi}=\frac{1}{2}^+$
with $4q$ partition $[31]_{orb}(0s^30p)[4]^{\sigma f}$ and seven
of $J^{\pi}=\frac{1}{2}^+$ with $4q$ partition
$[4]_{orb}(0s^30p)[31]^{\sigma f}$. Some modifications are made in
this further study, i.e., the orbital wave function is extended as
an expansion of $4$ different size harmonic oscillator forms;
three various forms (quadratic, linear and error function form) of
the color confinement potential are considered; the states with
$4q$ partition $[4]_{orb}(0s^30p)[31]^{\sigma f}$ are added, which
are unnegligible in the $J^{\pi}=\frac{1}{2}^+$ case and were not
considered in our previous paper, further the mixing between
configurations $[31]_{orb}(0s^30p)[4]^{\sigma f}$ and
$[4]_{orb}(0s^30p)[31]^{\sigma f}$ is also investigated. The
results show that the $T=0$ state is still always the lowest one
for both $J^{\pi}=\frac{1}{2}^-$ and $J^{\pi}=\frac{1}{2}^+$
states, and $J^{\pi}=\frac{1}{2}^-, T=0$ state is always lower
than that of $J^{\pi}=\frac{1}{2}^+$. All of these modifications
can only offer several tens to hundred MeV effect, and the
theoretical value of the lowest state is still about $245$ MeV
higher than the experimental mass of $\Theta^+$. It seems to be
difficult to get the calculated mass close to the observed one
with the reasonable parameters in the framework of the chiral
SU(3) quark model when the model space is chosen as a $5q$
cluster.

\end{abstract}

\vspace{0.3cm}

{\it PACS number(s):} 12.39.-x; 13.75.Jz; 21.45.+v


\vspace{1.5cm}

\section{Introduction}

Last year some Labs. reported that they observed a new resonance
state $\Theta$ with positive strangeness ${\cal{S}} = +1$
\cite{tna03,ale04}. The mass of this $\Theta$ particle is around
$M_{\Theta}=1540$ MeV and the upper limit of the width is about
$\Gamma_\Theta<25$ MeV. People suggested this exotic particle as a
pentaquark state because its strangeness quantum number is +1.
Although there are also several negative reports from some other
Labs. \cite{ale04,iab04}, it has motivated an enormous amount of
experimental and theoretical studies in the baryon physics area,
because if the existence of this exotic particle can be confirmed,
it will be the first multi-quark state people discovered. Actually
there have been a lots of theoretical works on the study of the
pentaquark baryons with various quark models, e.g. \cite{lyg03,
mkh03, eca03, fhu04, sac04}, and other approaches \cite{slz03,
ssa03, raa03, fjl04}, but its structure is still a challenging
problem.

It is well known that the non-perturbative quantum chromodynamics
(NPQCD) effect is very important in the light quark system, but up
to now there is no serious approach to really solve the NPQCD
problem. In this sense, people still need QCD-inspired models to
help in the low energy region. The constituent quark model
\cite{nis77, lyg96} is quite successful in explaining the baryon
spectrum, especially when the non-perturbative QCD effect is
considered by introducing the chiral field coupling \cite{lyg96,
pns97}, the Roper resonance and the low excited states of
$\Lambda$ can be explained simultaneously. At the same time, the
chiral SU(3) quark model can reasonably reproduce the binding
energy of deuteron, the nucleon-nucleon ($NN$) and the
kaon-nucleon ($KN$) scattering phase shifts of different partial
waves, and the hyperon-nucleon ($YN$) cross sections by the
resonating group method (RGM) calculations \cite{afa82, zyz77,
lrd03, fha04}. Inspired by these achievements, we try to extend
this model to study the $5$ quark system.

In our previous work \cite{fhu04}, we calculated the energies of
eight low configurations of the $5q$ system, four lowest
configurations of $J^{\pi}={\frac{1}{2}}^-$ with $4q$ partition
$[4]_{orb}(0s^4)[31]^{\sigma f}$ and four of
$J^{\pi}={\frac{1}{2}}^+$ with $4q$ partition
$[31]_{orb}(0s^30p)[4]^{\sigma f}$. But the results of the
adiabatic approximation calculation show that the mass of the
lowest $5q$ cluster state is about $150-300$ MeV higher than the
observed one of the $\Theta^+$ particle when the model parameters
are taken in the reasonable region.

The purpose of this paper is to do a further study on the $5q$
system with strangeness ${\cal{S}}=+1$ based on Ref.\cite{fhu04}.
Some modifications are made: (1) The orbital wave function is
extended as an expansion of the harmonic oscillator form with 4
different sizes $b_i(i=1-4)$ to improve the adiabatic
approximation. (2) Three various forms (quadratic, linear and
error function form) of the color confinement potential are
considered to examine the effects from the different confinement
potential. (3) Seven low lying $J^{\pi}={\frac{1}{2}}^+$ states of
partition $[4]_{orb}(0s^30p)[31]^{\sigma f}$ are added and the
mixing between configurations $[31]_{orb}(0s^30p)[4]^{\sigma f}$
and $[4]_{orb}(0s^30p)[31]^{\sigma f}$ is also studied. Meanwhile,
the parameters are chosen three different groups, one is fitted by
the $NN$ and $YN$ scattering experimental data \cite{zyz77, lrd03}
and the other two are fitted by the $KN$ scattering phase shifts
\cite{fha04}. The results show that with various groups of
parameters and different color confinement potentials, the $T=0$
state is still always the lowest one for both
$J^{\pi}={\frac{1}{2}}^-$ and $J^{\pi}={\frac{1}{2}}^+$ states,
and $J^{\pi}={\frac{1}{2}}^-$, $T=0$ state is always lower than
that of $J^{\pi}={\frac{1}{2}}^+$. In addition, the modification
of the adiabatic approximation and the confinement potential of
error function form can improve the calculated energy by several
tens MeV to the lowest state, respectively. When the parameters
are taken for the case by fitting the $KN$ phase shifts, the
energies of the system are much higher than those for the case by
fitting $NN$ and $YN$ data, because the $S$ wave interaction
between $K$ and $N$ is repulsive. Some of the
$J^{\pi}={\frac{1}{2}}^+$ states with 4q partition
$[4]_{orb}(0s^30p)[31]^{\sigma f}$ are lower than those with
partition of $[31]_{orb}(0s^30p)[4]^{\sigma f}$ and the mixing
between these two configurations can make the results $40-90$ MeV
lower. As a consequence, all of these modifications can offer
several tens to hundred MeV effect, and the calculated energy of
the lowest state is still about $245$ MeV higher than the
experimental mass of $\Theta$. It seems that when the model space
is chosen as $5q$ cluster, it is difficult to get the calculated
mass close to the observed one by using the chiral quark model
with the reasonable parameters.

The paper is arranged as follows. The theoretical framework of the
chiral SU(3) quark model and the determination of parameters are
briefly introduced in Section 2. The calculation results of
different confinement potentials with three groups of parameters
are listed and discussed in Section 3. Finally conclusions are
drawn in Section 4.

\section{Theoretical framework}
\subsection{The model}

As mentioned in Ref \cite{fhu04}, the Hamiltonian of the $5q$
system is written as

$$
    H=\sum_i{T_i}-T_G+\sum_{i<j=1}^{4}{V_{ij}}+\sum_{i=1}^4{V_{i5}},\eqno(1)
$$

\vskip .5cm \noindent where $\sum_i{T_i}-T_G$ is the kinetic
energy of the system, $V_{ij}(i,j=1-4)$ and $V_{i5}(i=1-4)$
represent the interactions between quark-quark ($q-q$) and
quark-antiquark ($q-\bar{q}$) respectively. In the chiral SU(3)
quark model, the quark-quark interaction includes three parts:
color confinement potential $V_{ij}^{conf}$, one gluon exchange
(OGE) interaction $V_{ij}^{OGE}$ and chiral field coupling induced
interaction $V_{ij}^{ch}$,

$$
   V_{ij}=V_{ij}^{conf}+V_{ij}^{OGE}+V_{ij}^{ch},\eqno(2)$$

\vskip .5cm \noindent where the confinement potential
$V_{ij}^{conf}$, which provides the non-perpurbative QCD effect in
the long distance, is taken as three different forms in this work,
i.e. quadratic, linear, and error function form. And the
expression of $V_{ij}^{OGE}$ is

$$
    V_{ij}^{OGE}=\frac{1}{4} g_i g_j \left(\lambda_i^c \cdot \lambda_j^c\right)
    \left\{\frac{1}{r_{ij}}-\frac{\pi}{2}
    \delta\left(\vec{r}_{ij}\right)
    \left(\frac{1}{m_{qi}^2}+\frac{1}{m_{qj}^2}+\frac{3}{4}\frac{1}{m_{qi}m_{qj}}
    \left(\vec{\sigma}_i \cdot \vec{\sigma}_j\right)\right)\right\} $$
$$
    +V_{\textit{tensor}}^{OGE}+V_{\vec{\textit{l}} \cdot
    \vec{\textit{s}}}^{OGE},~~~~~~~~~~~~~~~~~~~~~~~~~~~~~~~~~~~~~~~~~~~~~~~~~~\eqno(3)
$$

\vskip .5cm \noindent which governs the short-range perturbative
QCD behavior. $V_{ij}^{ch}$ represents the interactions from
chiral field couplings and describes the nonperturbative QCD
effect of the low-momentum medium-distance range, which can be
derived from the chiral-quark coupling interaction Lagrangian

$$
    {\cal{L}}_I^{ch}=-g_{ch} F(q^2) \bar{\psi} \left(\sum_{a=0}^8
    \sigma_a \lambda_a+i \sum_{a=0}^8 \pi_a \lambda_a
    \gamma_5\right)\psi,\eqno(4)
$$

\vskip .5cm \noindent where $\lambda_0$ is a unitary matrix,
$\sigma_0$,....,$\sigma_8$ are the scalar nonet fields and
$\pi_0$,....,$\pi_8$ the pseudoscalar nonet fields.
${\cal{L}}_I^{ch}$ is invariant under the infinitesimal chiral
SU(3)$_L \times$ SU(3)$_R$ transformation. Thus $V_{ij}^{ch}$ can
be expressed as

$$
    V_{ij}^{ch}=\sum_{a=0}^{8}V_{s_a}(\vec{r}_{ij})+
    \sum_{a=0}^{8}V_{ps_a}(\vec{r}_{ij}).\eqno(5)
$$

\vskip .5cm \noindent The expressions of $V_{s_a}(\vec{r}_{ij})$
and $V_{ps_a}(\vec{r}_{ij})$ can be found in Refs \cite{zyz77,
lrd03}.

The interaction between $q$ and $\bar{q}$ includes two parts:
direct interaction and annihilation part,

$$
   V_{q\bar{q}}=V_{q\bar{q}}^{dir}+V_{q\bar{q}}^{ann},\eqno(6)$$

$$
   V_{q\bar{q}}^{dir}=V_{q\bar{q}}^{conf}+V_{q\bar{q}}^{OGE}+V_{q\bar{q}}^{ch},\eqno(7)$$

\vskip .5cm\noindent with
$$
   V_{q\bar{q}}^{ch}(\vec{r})=\sum_i (-1)^{G_i}
   V_{qq}^{ch,i}(\vec{r}).\eqno(8)
$$

\vskip .5cm\noindent Here $(-1)^{G_i}$ describes the $G$ parity of
the $i$th meson. For the $\Theta$ particle case without vector
meson exchanges, $q\bar{q}$ can only annihilation into a $K$
meson, thus $V_{i5}^{ann}$ can be expressed as

$$
   V_{q\bar{q}}^{ann}=V_{ann}^{K},\eqno(9)$$

\vskip .5cm\noindent with
$$
   V_{ann}^K=C^{K}_{ann}
   \left(\frac{1-\vec{\sigma}_q \cdot
   \vec{\sigma}_{\bar{q}}}{2}\right)_{spin}
   \left(\frac{2+3\lambda_q \cdot \lambda_{\bar{q}}^{\ast}}{6}\right)_{color}
  \left(\frac{19}{9}+\frac{1}{6}\lambda_q \cdot
  \lambda_{\bar{q}}^{\ast}\right)_{flavor}\delta\left(\vec{r}_q-\vec{r}_{\bar{q}}\right),\eqno(10)
$$

\vskip .5cm\noindent where we treat $C^K_{ann}$ as a parameter and
adjust it to fit the mass of the $K$ meson.

In this work, calculations are carried on with three groups of
parameters. First, we take the parameters which can reasonably
reproduce the experimental data of $NN$ and $NY$ scattering (case
$\textrm{I}$) \cite{zyz77, lrd03}. By some special constraints,
the model parameters are fixed in the following way: the chiral
coupling constant $g_{ch}$ is fixed by
$$
    \frac{g_{ch}^2}{4\pi}=\left(\frac{3}{5}\right)^2
    \frac{g_{NN_{\pi}}^2}{4\pi} \frac{m_u^2}{M_N^2},\eqno(11)
$$
\vskip .5cm \noindent with $g_{NN_{\pi}}^2/4\pi=13.67$ taken as
the experimental value. The mass of the mesons are also adopted
the experimental values, except for the $\sigma$ meson, whose mass
is treated as an adjustable parameter. $\eta$, $\eta^{\prime}$
mesons are mixed by $\eta_0$, $\eta_8$,
$$
   \eta^{\prime}=\eta_8 \sin \theta^{PS}+\eta_0 \cos \theta^{PS},
$$
$$
   \eta=\eta_8 \cos \theta^{PS}-\eta_0 \sin \theta^{PS},\eqno(12)
$$
\vskip .5cm\noindent with the mixing angle $\theta^{PS}$ taken to
be the usual value $-23^{\circ}$.

The one gluon exchange coupling constants $g_u$ and $g_s$ can be
determined by the mass splits between $N$, $\Delta$ and $\Sigma$,
$\Lambda$ respectively. The confinement strengths $a_{uu}^c$,
$a_{us}^{c}$, and $a_{ss}^c$ are fixed by the stability conditions
of $N$, $\Lambda$, and $\Xi$, and the zero point energies
$a_{uu}^{c0}$, $a_{us}^{c0}$, and $a_{ss}^{c0}$ by fitting the
masses of $N$, $\Sigma$, and $\overline{\Xi+\Omega}$,
respectively.

Another two groups of parameters are fitted by $KN$ scattering
\cite{fha04}, in which the scalar meson mixing between the flavor
singlet and octet mesons is considered, i.e. $\sigma$ and
$\epsilon$ mesons are mixed by $\sigma_0$ and $\sigma_8$,
$$
    \sigma=\sigma_8 \sin \theta^S+\sigma_0 \cos\theta^S,
$$
$$
    \epsilon=\sigma_8 \cos \theta^S-\sigma_0 \sin\theta^S.\eqno(13)
$$
\noindent The mixing angle $\theta^S$ is an open problem because
the structure of the $\sigma$ meson is unclear and controversial.
We adopt two possible values by which we can get reasonable $KN$
phase shifts, one is ideally mixing $\theta^S=35.264^{\circ}$
(case \textrm{II}), which means that $\sigma$ only act on the
$u(d)$ quark, and $\epsilon$ on the $s$ quark, the other is
$\theta^S=-18^{\circ}$ (case \textrm{III}), which is provided by
Dai and Wu based on their recent investigation \cite{ybd03}.

The three sets of model parameters are tabulated in Table
\ref{para}, the first column (case I) is for the case fitted by
$NN$ and $YN$ scattering, and the second (case II) and third (case
III) columns for the case fitted by $KN$ scattering.

{\small
\begin{table}[p]
\caption{\label{para} Model parameters. The meson masses and the
cut-off masses: $m_{\sigma^{\prime}}= 980$ MeV, $m_{\kappa}=980$
MeV, $m_{\epsilon}=980$ MeV, $m_{\pi}=138$ MeV, $m_{K}=495$ MeV,
$m_{\eta}=549$ MeV, $m_{\eta^{\prime}}=957$ MeV, $\Lambda=1100$
MeV.} \setlength{\tabcolsep}{2.3mm}
\begin{center}
\begin{tabular}{ccccccc}
\hline \hline
 &&For $NN,YN$ cases (I)&&\multicolumn{3}{c}{For $KN$ case}\\
\cline{5-7}
&&&&$\theta^S=35.264^{\circ}$(II)&&$\theta^S=-18^{\circ}$(III)\\
\hline
$b_u$(fm)&&0.5&&0.5&&0.5\\
$m_u$(MeV)&&313&&313&&313\\
$m_s$(MeV)&&470&&470&&470\\
$g_u$&&0.886&&0.886&&0.886\\
$g_s$&&0.755&&0.755&&0.755\\
$m_{\sigma}$(MeV)&&595&&675&&675\\
$\ast~a_{uu}^c$(MeV/fm$^2$)&&48.1&&52.4&&55.2\\
$\ast~a_{us}^c$(MeV/fm$^2$)&&60.7&&72.3&&68.4\\
$\ast~a_{uu}^{c0}$(MeV)&&$-$43.6&&$-$50.4&&$-$55.1\\
$\ast~a_{us}^{c0}$(MeV)&&$-$38.3&&$-$54.2&&$-$48.7\\
$\diamond~a_{uu}^c$(MeV/fm)&&86.6&&95.1&&100.6\\
$\diamond~a_{us}^c$(MeV/fm)&&101.0&&121.7&&114.7\\
$\diamond~a_{uu}^{c0}$(MeV)&&$-$75.6&&$-$86.0&&$-$93.1\\
$\diamond~a_{us}^{c0}$(MeV)&&$-$72.3&&$-$96.5&&$-$88.2\\
$\star~a_{uu}^c$(MeV)&&194.4&&213.3&&225.7\\
$\star~a_{us}^c$(MeV)&&218.2&&262.9&&247.8\\
$\star~a_{uu}^{c0}$(MeV)&&$-$87.4&&$-$99.0&&$-$106.8\\
$\star~a_{us}^{c0}$(MeV)&&$-$82.8&&$-$109.1&&$-$100.0\\
\hline \hline
\end{tabular}
\end{center}
Here $a_{uu}^c$, $a_{us}^c$, $a_{uu}^{c0}$, $a_{uu}^{c0}$ with
symbol '$\ast$' are for the quadratic form color confinement,
those with '$\diamond$' are for the linear form confinement, and
those with '$\star$' are for the error function form confinement.
\end{table}}

\subsection{The configurations}

For the $5q$ system with ${\cal{S}}=+1$, there are a lot of
different configurations. In the actually calculating process, we
have to choose some lower configurations. Using this model, we
considered 15 states of the $5q$ system with strangeness
${\cal{S}}=+1$: four lowest $J^{\pi}=\frac{1}{2}^-$ states with
$4q$ partition $[4]_{orb}(0s)^4$ \vskip .3cm
$[4]_{orb}[31]_{ts=01}^{\sigma f}
\bar{s},LST=0\frac{1}{2}0,J^{\pi}=\frac{1}{2}^-$,~~~~
$[4]_{orb}[31]_{ts=10}^{\sigma f}
\bar{s},LST=0\frac{1}{2}1,J^{\pi}=\frac{1}{2}^-$,\vskip .3cm

$[4]_{orb}[31]_{ts=11}^{\sigma f}
\bar{s},LST=0\frac{1}{2}1,J^{\pi}=\frac{1}{2}^-$,~~~~
$[4]_{orb}[31]_{ts=21}^{\sigma f}
\bar{s},LST=0\frac{1}{2}2,J^{\pi}=\frac{1}{2}^-$,

\vskip .3cm

\noindent and four configurations of $J^{\pi}=\frac{1}{2}^+$ with
$[31]_{orb}(0s)^3(0p)$

\vskip .3cm $[31]_{orb}[4]_{ts=00}^{\sigma f}
\bar{s},LST=1\frac{1}{2}0,J^{\pi}=\frac{1}{2}^+$,~~~~
$[31]_{orb}[4]_{ts=11}^{\sigma f}
\bar{s},LST=1\frac{1}{2}1,J^{\pi}=\frac{1}{2}^+$,\vskip .3cm
$[31]_{orb}[4]_{ts=11}^{\sigma f}
\bar{s},LST=1\frac{3}{2}1,J^{\pi}=\frac{1}{2}^+$,~~~~
$[31]_{orb}[4]_{ts=22}^{\sigma f}
\bar{s},LST=1\frac{3}{2}2,J^{\pi}=\frac{1}{2}^+$. \vskip .3cm

\noindent Here the symbols $[f]_{orb}$ and $[f^{\prime}]^{\sigma
f}$ are the partitions of orbital space and flavor-spin space
respectively; $[4]_{orb}$ represents the total symmetric state in
the orbital space, where four quarks are all in $(0s)$ state; and
$[31]_{orb}$ is the orbital space partition of $(0s)^3(0p)$.
Symbol $t$ and $s$ denote the isospin and spin of the four quark
part; after coupling with the fifth quark $\bar{s}$, the total
orbital angular momentum, spin and isospin of the five quark
system are expressed as $LST$. The color part is $(10)_c$ of $4q$
and $(01)_c$ of the anti-quark $\bar{s}$ respectively, here we
omitted them in the expressions.

Meanwhile, there are some other $J^{\pi}=\frac{1}{2}^+$ states (we
did not consider them in our previous work \cite{fhu04}) in which
the $\bar{s}$ can be in the $(0p)$ state. The wave functions of
these states should be orthogonal to the excited states of the
center of mass motion. The expression of the wave function with
various $ts$ and $LST$, $\Psi^{LST}_{ts}(5q)$, is given as
following

$$
    \Psi^{LST}_{ts}(5q)=\sqrt{\frac{m_s}{4 m_u + m_s}}~
        \left(\Psi_{4q}((0s^30p)[4]_{orb}[31]^{\sigma
        f}_{ts}[211]_c)\Phi_{0s}(\bar{s})\right)_{LST}$$
$$  ~~~~~~~~~~~  -\sqrt{\frac{4 m_u}{4 m_u + m_s}}~
    \left(\Psi_{4q}((0s^4)[4]_{orb}[31]^{\sigma
f}_{ts}[211]_c)\Phi_{0p}(\bar{s})\right)_{LST}.\eqno(14)
$$

\vskip .5cm\noindent Where in the first term, the $4q$ orbital
wave function is $(0s)^3(0p)$ with partition $[4]_{orb}$ and
$\bar{s}$ is in $(0s)$, while in the second term, the $4q$ orbital
wave function is $(0s)^4$ with partition $[4]_{orb}$ and $\bar{s}$
is in $(0p)$. $[31]^{\sigma f}_{ts}$ is the $4q$ partition of
spin-flavor space. The total orbital angular momentum, spin and
isospin of the five quark system are $LST$. We considered seven
low states of $\Psi^{LST}_{ts}(5q)$, they are

\vskip .3cm $[4]_{orb}[31]_{ts=01}^{\sigma f}
\bar{s},LST=1\frac{1}{2}0,J^{\pi}=\frac{1}{2}^+$,~~~~~
$[4]_{orb}[31]_{ts=10}^{\sigma f}
\bar{s},LST=1\frac{1}{2}1,J^{\pi}=\frac{1}{2}^+$, \vskip .3cm
$[4]_{orb}[31]_{ts=11}^{\sigma f}
\bar{s},LST=1\frac{1}{2}1,J^{\pi}=\frac{1}{2}^+$,~~~~~
$[4]_{orb}[31]_{ts=21}^{\sigma f}
\bar{s},LST=1\frac{1}{2}2,J^{\pi}=\frac{1}{2}^+$,\vskip .3cm
$[4]_{orb}[31]_{ts=01}^{\sigma f}
\bar{s},LST=1\frac{3}{2}0,J^{\pi}=\frac{1}{2}^+$,~~~~~
$[4]_{orb}[31]_{ts=11}^{\sigma f}
\bar{s},LST=1\frac{3}{2}1,J^{\pi}=\frac{1}{2}^+$,\vskip .3cm
$[4]_{orb}[31]_{ts=21}^{\sigma f}
\bar{s},LST=1\frac{3}{2}2,J^{\pi}=\frac{1}{2}^+$.\\

In order to improve the adiabatic approximation, in this work the
so-called "breath model" is taken to carry on the energy
calculation. The trail wave function can be written as an
expansion of the $5q$ states with several different harmonic
oscillator frequency $\omega_i$,

$$
   \Psi_{5q}=\sum_{i}^{n} \alpha_i\Phi_{5q}(b_i).\eqno(15)
$$

\noindent Where $(b_i)^2=\frac{1}{m\omega_i}$. Using the
fractional parentage coefficient (f.p.) technique, we can easily
write down the wave functions of the above 15 states and by
solving the Schr\"{o}dinger equation the energies of these states
are obtained.

\section{Results and discussions}

We calculate energies of fifteen low configurations, four lowest
configurations of $J^{\pi}=\frac{1}{2}^-$ with $4q$ partition
$[4]_{orb}(0s^4)[31]^{\sigma f}$, four of $J^{\pi}=\frac{1}{2}^+$
with $4q$ partition $[31]_{orb}(0s^30p)[4]^{\sigma f}$ and seven
low configurations of $J^{\pi}=\frac{1}{2}^+$ with $4q$ partition
$[4]_{orb}(0s^30p)[31]^{\sigma f}$. Comparing with our previous
work \cite{fhu04}, some modifications are made in this work. The
trial wave function is taken as an expansion of harmonic
oscillator wave functions with four different size parameter
$b_i(i=1-4)$ to improve the adiabatic approximation. Three
different forms (quadratic, linear and error function form) of the
confinement potential are considered to study the effects from
various confinement potentials. More low lying states of
$J^{\pi}={\frac{1}{2}}^+$ with $4q$ partition $[4]_{orb}(0s^30p)$
and the mixing between some configurations are considered. At the
same time, parameters are chosen three different groups as listed
in Table \ref{para}. The results are given in Table
\ref{results1}. In this table, case $\textrm{I}$ means the
parameters are fitted by $NN$ and $YN$ scattering, while case
$\textrm{II}$ by $KN$ scattering with $\theta^S=35^{\circ}$ and
case $\textrm{III}$ by $KN$ scattering with
$\theta^S=-18^{\circ}$. And $r^2$, $r$ and $erf$ represent the
confinement potential is adopted as quadratic, linear and error
function form respectively.

{\small
\begin{table}[p]
\caption{\label{results1}Energies(in MeV) of fifteen
configurations with 4 different size $b_i(i=1-4)$. }
\renewcommand{\arraystretch}{1.15}
\setlength{\tabcolsep}{1.7mm}
\begin{center}
\begin{tabular}{ccccccccccccc} \hline \hline
configurations&&\multicolumn{3}{c}{case I}&&\multicolumn{3}{c}{case II}&&\multicolumn{3}{c}{case III}\\
\hline
 & & $r^2$ & $r$ & $erf$ & & $r^2$ & $r$ & $erf$ & & $r^2$ & $r$ & $erf$\\
\raisebox{2.3ex}[0pt]{$J^{\pi}=\frac{1}{2}^-$}&& &(MeV)& &&
&(MeV)& && &(MeV)& \\
\hline
$([4]_{orb}[31]_{ts=01}^{\sigma f}\bar{s})_{0\frac{1}{2}0}$ && 1799 & 1790 & 1785 && 1901 & 1890 & 1884 && 1897 & 1886 & 1880\\

$([4]_{orb}[31]_{ts=10}^{\sigma f}\bar{s})_{0\frac{1}{2}1}$ && 2085 & 2061 & 2044 && 2155 & 2127 & 2107 && 2156 & 2128 & 2108\\

$([4]_{orb}[31]_{ts=11}^{\sigma f}\bar{s})_{0\frac{1}{2}1}$ && 2150 & 2120 & 2098 && 2219 & 2184 & 2159 && 2220 & 2185 & 2160\\

$([4]_{orb}[31]_{ts=21}^{\sigma f}\bar{s})_{0\frac{1}{2}2}$ && 2353 & 2294 & 2251 && 2417 & 2352 & 2305 && 2420 & 2355 & 2308\\
\hline
 && $r^2$ & $r$ & $erf$ && $r^2$ & $r$ & $erf$ && $r^2$ & $r$ & $erf$\\
\raisebox{2.3ex}[0pt]{$J^{\pi}=\frac{1}{2}^+$}&& &(MeV)& && &(MeV)& && &(MeV)& \\
\hline

$([31]_{orb}[4]_{ts=00}^{\sigma f}\bar{s})_{1\frac{1}{2}0}$ && 2153 & 2129 & 2114 && 2221 & 2193 & 2174 && 2206 & 2179 & 2162\\

$([31]_{orb}[4]_{ts=11}^{\sigma f}\bar{s})_{1\frac{1}{2}1}$ && 2203 & 2173 & 2154 && 2274 & 2239 & 2216 && 2257 & 2224 & 2202\\

$([31]_{orb}[4]_{ts=11}^{\sigma f}\bar{s})_{1\frac{3}{2}1}$ && 2257 & 2221 & 2198 && 2323 & 2282 & 2255 && 2308 & 2269 & 2243\\

$([31]_{orb}[4]_{ts=22}^{\sigma f}\bar{s})_{1\frac{3}{2}2}$ && 2337 & 2291 & 2257 && 2409 & 2354 & 2317 && 2391 & 2339 & 2304\\
\hline
 && $r^2$ & $r$ & $erf$ && $r^2$ & $r$ & $erf$ && $r^2$ & $r$ & $erf$\\
\raisebox{2.3ex}[0pt]{$J^{\pi}=\frac{1}{2}^+$}&& &(MeV)& && &(MeV)& && &(MeV)& \\
\hline
$([4]_{orb}[31]_{ts=01}^{\sigma f}\bar{s})_{1\frac{1}{2}0}$ && 2071 & 1958 & 1917 && 2110 & 1975 & 1924 && 2123 & 1994 & 1945\\

$([4]_{orb}[31]_{ts=10}^{\sigma f}\bar{s})_{1\frac{1}{2}1}$ && 2239 & 2105 & 2048 && 2267 & 2110 & 2043 && 2283 & 2132 & 2066\\

$([4]_{orb}[31]_{ts=11}^{\sigma f}\bar{s})_{1\frac{1}{2}1}$ && 2294 & 2152 & 2089 && 2322 & 2157 & 2083 && 2337 & 2178 & 2107\\

$([4]_{orb}[31]_{ts=21}^{\sigma f}\bar{s})_{1\frac{1}{2}2}$ && 2458 & 2285 & 2198 && 2485 & 2289 & 2192 && 2501 & 2311 & 2216\\

$([4]_{orb}[31]_{ts=01}^{\sigma f}\bar{s})_{1\frac{3}{2}0}$ && 2275 & 2138 & 2077 && 2300 & 2140 & 2070 && 2317 & 2162 & 2094\\

$([4]_{orb}[31]_{ts=11}^{\sigma f}\bar{s})_{1\frac{3}{2}1}$ && 2270 & 2132 & 2071 && 2298 & 2137 & 2066 && 2313 & 2158 & 2090\\

$([4]_{orb}[31]_{ts=21}^{\sigma f}\bar{s})_{1\frac{3}{2}2}$ && 2253 & 2200 & 2127 && 2383 & 2205 & 2133 && 2398 & 2227 & 2146\\
\hline \hline
\end{tabular}
\end{center}
Configurations are express as: $([f]_{orb}[f^{\prime}]^{\sigma
f}_{ts}\bar{s})_{LST}$, in which $LST$ are  the total orbital
angular momentum, spin and isospin of the five quark system.
\end{table}}

{\small
\begin{table}[p]
\caption{\label{compare1}Comparison between the energies with
$1~b$ (adiabatic approximation) and $4~b_i$ of three lowest
configurations of $J^{\pi}=\frac{1}{2}^-$ and
$J^{\pi}=\frac{1}{2}^+$ by taking the parameters of case I, which
are fitted by the $NN$ and $YN$ scattering experimental data.}
\renewcommand{\arraystretch}{1.15}
\setlength{\tabcolsep}{3.25mm}
\begin{center}
\begin{tabular}{ccccccccc} \hline \hline
 & & $r^2$ & $r$ & $erf$ & & $r^2$ & $r$ & $erf$  \\
 & &\multicolumn{3}{c}{$1~b^{\ast}~~~~$(MeV)}& &
\multicolumn{3}{c}{$4~b~~~~~$(MeV)}  \\
\hline
$([4]_{orb}[31]_{ts=01}^{\sigma f}\bar{s})_{0\frac{1}{2}0}$, $J^{\pi}=\frac{1}{2}^-$ & & 1821 & 1820 & 1820 & & 1799 & 1790 & 1785 \\
$([31]_{orb}[4]_{ts=00}^{\sigma f}\bar{s})_{1\frac{1}{2}0}$, $J^{\pi}=\frac{1}{2}^+$ & & 2162 & 2145 & 2135 & & 2153 & 2129 & 2114  \\
$([4]_{orb}[31]_{ts=01}^{\sigma f}\bar{s})_{1\frac{1}{2}0}$, $J^{\pi}=\frac{1}{2}^+$ & & 2079 & 1973 & 1936 & & 2071 & 1958 & 1917 \\
\hline \hline
\end{tabular}
\end{center}
$\ast$ Here the energies with $1b$ are a little bit higher than
those in Ref. \cite{fhu04}. This is because that the annihilation
to $K^{\ast}$ is not considered in our present calculation, but it
is involved in Ref. \cite{fhu04}.
\end{table}}

{\small
\begin{table}[p]
\caption{\label{results2}Comparison between the energies without
and with configuration mixing between two
$J^{\pi}={\frac{1}{2}}^+$ states by taking the parameters of case
I. }
\renewcommand{\arraystretch}{1.15}
\setlength{\tabcolsep}{3.55mm}
\begin{center}
\begin{tabular}{ccccccccccc} \hline \hline
 & & &\multicolumn{3}{c}{without mixing}& & &\multicolumn{3}{c}{with mixing}\\
\hline
 & & & $r^2$ & $r$ & $erf$ & & & $r^2$ & $r$ & $erf$ \\
\raisebox{2.3ex}[0pt]{$J^{\pi}=\frac{1}{2}^+$}& & & &(MeV)& & & &
&(MeV)&  \\
\hline
       & & & 2203 & 2152 & 2089 & & & 2165 & 2054 & 2018 \\
\raisebox{2.3ex}[0pt]{$LST=1\frac{1}{2}1$}& & & 2294 & 2173 & 2154 & & & 2331 & 2271 & 2225 \\
\hline
      & & & 2257 & 2132 & 2071 & & & 2220 & 2082 & 2032 \\
\raisebox{2.3ex}[0pt]{$LST=1\frac{3}{2}1$}  & & & 2270 & 2221 & 2198 & & & 2306  & 2271 & 2237\\
\hline \hline
\end{tabular}
\end{center}
  In the column with the configuration mixing, the third row gives two groups of results with considering
the mixing between $([31]_{orb}[4]_{ts=11}^{\sigma
f}\bar{s})_{1\frac{1}{2}1}$ and $([4]_{orb}[31]_{ts=11}^{\sigma
f}\bar{s})_{1\frac{1}{2}1}$, and the forth row gives those with
considering the mixing between $([31]_{orb}[4]_{ts=11}^{\sigma
f}\bar{s})_{1\frac{3}{2}1}$ and $([4]_{orb}[31]_{ts=11}^{\sigma
f}\bar{s})_{1\frac{3}{2}1}$.
\end{table}}

{\small
\begin{table}[htb]
\caption{\label{compare2}Comparison between the energies with
$b_u=0.5$ fm and $b_u=0.6$ fm
}
\renewcommand{\arraystretch}{1.15}
\setlength{\tabcolsep}{3.35mm}
\begin{center}
\begin{tabular}{ccccccccc} \hline \hline
 & & $r^2$ & $r$ & $erf$ & & $r^2$ & $r$ & $erf$  \\
 & &\multicolumn{3}{c}{$b_u=0.5$ fm~~(MeV)}& &
\multicolumn{3}{c}{$b_u=0.6$ fm~~(MeV)}  \\
\hline
$([4]_{orb}[31]_{ts=01}^{\sigma f}\bar{s})_{0\frac{1}{2}0}$, $J^{\pi}=\frac{1}{2}^-$ & & 1799 & 1790 & 1785 & & 1668 & 1666 & 1665 \\
$([31]_{orb}[4]_{ts=00}^{\sigma f}\bar{s})_{1\frac{1}{2}0}$, $J^{\pi}=\frac{1}{2}^+$ & & 2153 & 2129 & 2114 & & 1976 & 1968 & 1963  \\
$([4]_{orb}[31]_{ts=01}^{\sigma f}\bar{s})_{1\frac{1}{2}0}$, $J^{\pi}=\frac{1}{2}^+$ & & 2071 & 1958 & 1917 & & 1887 & 1878 & 1875 \\
\hline \hline
\end{tabular}
\end{center}
\end{table}}

From Table \ref{results1}, one can see that: (1) the isoscalar
state $T=0$ is always the lowest state both in
$J^{\pi}=\frac{1}{2}^-$ and in $J^{\pi}=\frac{1}{2}^+$ cases, and
$[4]_{orb}[31]_{ts=01}^{\sigma f}\bar{s}$, $LST=0\frac{1}{2}0$,
$J^{\pi}=\frac{1}{2}^-$ is the lowest one among all the states.
For two sets of $J^{\pi}=\frac{1}{2}^+$ states, some
configurations with $4q$ partition $[4]_{orb}$, especially
($[4]_{orb}[31]_{ts=01}^{\sigma f}\bar{s}$, $LST=1\frac{1}{2}0$,
$J^{\pi}=\frac{1}{2}^+$), are about $100-200$ MeV lower than those
of $[31]_{orb}$ in various cases. It means that in
$J^{\pi}=\frac{1}{2}^+$ state, the configurations with $4q$
partition $[4]_{orb}$ can not be neglected. (2) The effect of
various confinement potentials is about several tens MeV. The
energies obtained from the error function confinement potential
are the lowest, and the influence is larger to the high energy
configurations. (3) The energies of case $\textrm{I}$ are lower
than those of the other two cases, this is obvious, because in the
$NN$ case the $S$ wave interaction is attraction, while in the
$KN$ case it is repulsive.

The comparison between the results of adiabatic approximation (the
wave function is taken as harmonic oscillator function with $1b$)
and those of the "breath model" (the wave function is treated as
an expansion of the harmonic oscillator functions with 4 different
frequency $\omega$) is shown in Table \ref{compare1}. Here the
parameters are taken as case $\textrm{I}$, which is fitted by $NN$
and $YN$ scattering data. The results show that this modification
can only reduce the energies about $10-30$ MeV for different
configurations. This means that even the trial wave function is
modified, the energy of the $5q$ system can not be improved a lot.

In the case of $J^{\pi}=\frac{1}{2}^+$,
$([31]_{orb}[4]_{ts=11}^{\sigma f}\bar{s})_{1\frac{1}{2}1}$ and
$([4]_{orb}[31]_{ts=11}^{\sigma f}\bar{s})_{1\frac{1}{2}1}$  as
well as $([31]_{orb}[4]_{ts=11}^{\sigma
f}\bar{s})_{1\frac{3}{2}1}$ and $([4]_{orb}[31]_{ts=11}^{\sigma
f}\bar{s})_{1\frac{3}{2}1}$ have the same quantum numbers. The
configuration mixing has to be considered for these two cases. The
results comparing with those without configuration mixing are
given in Table \ref{results2}. From this table, one can see that
the configurations mixing effect is not very small, it can make
energies about $40-90$ MeV lower.

We also try to adjust the size parameter $b_u$ to be larger to see
the influence. As an example, the results of lower configurations
with $b_u=0.6$ fm in the chiral SU(3) quark model are given
compared with the former results with $b_u=0.5$ fm in Table
\ref{compare2}. In this case, the energies of all states become
smaller, caused by the kinetic energy of the system is reduced for
larger $b_u$. But the lowest energy $1665$ MeV is still about
$125$ MeV higher than the experimental mass of the observed
$\Theta$.

In addition, we notice that recently Sachiko Takeuchi $et~al.$
\cite{sac04} investigated $uudd\bar{s}$ pentaquarks by employing
their quark models with the meson exchange and the effective gluon
exchange as $qq$ and $q\bar{q}$ interactions, and dynamically
solved the system by taking two quarks as a diquark-like $qq$
correlation. Their calculated value of the lowest configuration
was $1947-2144$ MeV and the low mass close to the observed one
could not be obtained, which was similar to our results. The main
difference is that in Ref. \cite{sac04} there were parameter sets
where the mass of the lowest positive-parity states became lower
than that of the negative-parity states. However, in our work,
where the model parameters are fitted by the $NN$,$YN$ and $KN$
scattering phase shifts and the $q\bar q$ $s$-channel interactions
are fitted by the mass of the kaon meson, the results show that
the $J^\pi=\frac{1}{2}^-$, $T=0$ state is always the lowest one.
As discussed in our previous work \cite{fhu04}, if we omitted the
interactions between $4q$ and $\bar q$, then the state with
positive parity can be lower than that with negative parity, in
agreement with what is claimed by Takeuchi and Shimizu
\cite{sac04}, and Stancu and Riska \cite{lyg03}. This means that
how to treat the annihilation interactions reasonably is very
important in the calculation.

Meanwhile, it is important to investigate the narrow width as well
as the low mass of the $\Theta$ particle. But in our present work,
we concentrate our study on the masses of some low-lying $5q$
configurations with various quantum numbers. Since the coupling
between $5q$ configuration states and continuum baryon-meson
states such as $KN$ is not included, we can only get some
qualitative information about the width from the wave functions.
For example, by using the group theory method\cite{fi96} the
lowest state of $J^{\pi}=\frac{1}{2}^-$ and $T=0$,
$\Psi([4]_{orb}(0s)^4[31]_{ts=01}^{\sigma
f}\bar{s}(0s))_{LST=0\frac{1}{2}0,(00)_c}$, can be expanded as:
$$
\Psi([4]_{orb}(0s)^4[31]_{ts=01}^{\sigma
f}\bar{s}(0s))_{LST=0\frac{1}{2}0,(00)_c}=~~~~~~~~~~~~~~~~~~~~~~~~~~~~~~$$
$$-\frac{1}{2}~(\Phi_{123})_{(s=\frac{1}{2})(t=\frac{1}{2})(11)_c}(\Phi_{4\bar{5}})_{(s=0)(11)_c}
$$
$$
          -\frac{1}{2}\sqrt{\frac{1}{3}}~(\Phi_{123})_{(s=\frac{1}{2})(t=\frac{1}{2})(11)_c}(\Phi_{4\bar{5}})_{(s=1)(11)_c}$$
$$
+\sqrt{\frac{1}{3}}~(\Phi_{123})_{(s=\frac{3}{2})(t=\frac{1}{2})(11)_c}(\Phi_{4\bar{5}})_{(s=1)(11)_c}$$
$$
+\frac{1}{2}~(\Phi_{123})_{(s=\frac{1}{2})(t=\frac{1}{2})(00)_c}(\Phi_{4\bar{5}})_{(s=0)(00)_c}$$
$$
+\frac{1}{2}\sqrt{\frac{1}{3}}~(\Phi_{123})_{(s=\frac{1}{2})(t=\frac{1}{2})(00)_c}(\Phi_{4\bar{5}})_{(s=1)(00)_c},\eqno(16)
$$

\noindent where the wave function is antisymmetrized and all terms
are orthogonal and linearly independent each other. In the above
expression (Eq.(16)), the fourth term is the component of $S$-wave
$KN$, whose probability is $(\frac{1}{2})^2=\frac{1}{4}$. Such a
large $(25\%)$ $KN$ component would consequently make the width of
this state to be quite large. In this sense the $5q$ cluster
$\frac{1}{2}^+$ state can be expected to have a smaller width.
However in our present work, when the model parameters are fitted
by the $NN$,$YN$ and $KN$ scattering phase shifts, the results
show that the masses of the positive-parity states are higher than
those of the negative-parity states, also much higher than the
experimental value. In this framework, it is difficult to
understand both the low mass and the narrow width of the $\Theta$.
But up to now, the existence of the $\Theta$ particle is still a
controversial problem, and some high-statistics experimental
collaborations showed the negative results. From our results, we
can just say that if the $\Theta$ particle do exist, it can not be
explained as a five-constituent-quark cluster, and its structure
should be understood from other mechanism.

\section{Conclusions}

The structures of $5q$ cluster states with ${\cal{S}}=+1$ are
further studied based on the $5q$ cluster configurations studied
in the SU(3) chiral quark model. Fifteen low configurations and
the mixing between some configurations are considered. In the
calculation, the trial wave function is taken as an expansion of
harmonic oscillator wave functions with four different size
parameters. The effect of various color confinement potential is
examined by taken three different forms (quadratic, linear and
error function forms). With various groups of parameters and
different color confinement potentials, the isoscalar state $T=0$
is always the lowest state both in $J^{\pi}=\frac{1}{2}^-$ and in
$J^{\pi}=\frac{1}{2}^+$ cases. And $J^{\pi}=\frac{1}{2}^-$, $T=0$
is the lowest one among all of these configurations. The
modification of the adiabatic approximation and the confinement
potential of error function form can improve the calculated energy
by several tens MeV to the lowest state, respectively. Since the
$S$ wave interaction between $K$ and $N$ is repulsive, when the
parameters are taken for the case by fitting the $KN$ phase
shifts, the energies of the system are much higher than those for
the case by fitting $NN$ and $YN$ data. The configurations mixing
of $J^{\pi}={\frac{1}{2}}^+$ states can make the calculated mass
$40-90$ MeV shifted. If we adjust the size parameter larger,
$b_u=0.6$ fm, the energy of the lowest configuration will be
$1665$ MeV. All of these modifications can only offer several tens
to hundred MeV improvement, and the calculated value of the lowest
state is still about $125$ MeV higher than the experimental mass
of $\Theta$. It seems that when the model space is chosen as $5q$
cluster, it is difficult to get the calculated mass close to the
observed one in the framework of the chiral quark model with the
reasonable parameters. Though the existence of the $\Theta$
particle is still an open problem, this work just presents that it
can not be regarded as a five-constituent-quark cluster, if it do
exist, its structure should be explained by other mechanism.

\end{document}